\documentclass[twocolumn,showpacs,preprintnumbers,amsmath,amssymb,prb]{revtex4-1}
\usepackage{graphicx}% Include figure files
\usepackage{dcolumn}% Align table columns on decimal point
\usepackage{bm}% bold math
\usepackage{color}
\usepackage{physics}

\renewcommand{\d}{{\bm{\delta}}}
\newcommand{\e}{{\bm e}}

\newcommand{\p}{{\bf p}}

\renewcommand{\k}{{\bm{k}}}
\newcommand{\q}{{\bm{q}}}
\renewcommand{\r}{{\bm{r}}}

\def\lsim{\lower.35em\hbox{$\stackrel{\textstyle<}{\textstyle\sim}$}}
\def\gsim{\lower.35em\hbox{$\stackrel{\textstyle>}{\textstyle\sim}$}}

\begin{document}

\title{Marginal Fermi liquid in twisted bilayer graphene}

\author{J. Gonz\'alez$^{1}$ and T. Stauber$^{2}$}

\affiliation{
$^{1}$ Instituto de Estructura de la Materia, CSIC, E-28006 Madrid, Spain\\
$^{2}$ Materials Science Factory,
Instituto de Ciencia de Materiales de Madrid, CSIC, E-28049 Madrid, Spain}
\date{\today}

\begin{abstract}
Linear resistivity at low temperatures is a prominent feature of high-T$_c$ superconductors which has also been found recently in twisted bilayer graphene. We show that due to an extended van Hove singularity (vHS), the $T$-linear resistivity can be obtained from a microscopic tight-binding model for filling factors close to the vHS. The linear behavior is shown to be related to the linear energy dependence of the electron quasiparticle decay rate which implies the low-energy logarithmic attenuation of the quasiparticle weight. These are distinctive features of a marginal Fermi liquid, which we also see reflected in the respective low-temperature logarithmic corrections of the heat capacity and the thermal conductivity, leading to the consequent violation of the Wiedemann-Franz law. We also show that there is a crossover at $T \sim 6$ K from the marginal Fermi liquid regime to a regime dominated by excitations on the Dirac cone right above the vHS that also yields a linear resistivity albeit with smaller slope, in agreement with experimental observations.

\end{abstract}

\maketitle

{\it Introduction.} 
The discovery of a correlated insulating\cite{Cao18a} and superconducting\cite{Cao18b} state in magic angle twisted bilayer graphene (TBG) has stimulated great interest, both from the theoretical\cite{Xu18,Volovik18,Yuan18,Po18,Roy18,Guo18,Dodaro18,Baskaran18,Liu18,Slagle18,Peltonen18,Kennes18,Koshino18,Kang18,Isobe18,You18,Wu18b,Zhang18,Pizarro18,Pal18,Ochi18,Thomson18,Carr18,Guinea18,Zou18} as well as from the experimental side.\cite{Yankowitz19,Choi19,Sharpe19,Moriyama19,Jiang19,Xie19,Kerelsky19} 
One striking and astonishing result of the initial experiments was the similarity of the phase diagram to the one of high-$T_c$ superconductors.\cite{Micnas90,Damascelli03,Lee06,Stewart11} This analogy has further been manifested by the observation of a strange metal regime with its linear temperature dependence of the resistivity.\cite{Cao19,Polshyn19} 

The nature of the strange metal phase in TBG is highly controversial. It is frequently assumed that electron-phonon interactions could be responsible for the anomalous behavior of the resistivity, but at the same time acknowledging that phonons cannot account for the $T$-linear dependence down to the lowest temperatures reached in the experiments ($\sim 0.5$ K), in particular below the Debye frequency scale (in the case of optical phonons) or below the Bloch-Gr\"uneisen temperature (in the case of acoustic phonons). The proposal derived in this Letter solves this puzzle, presenting a consistent explanation that is purely based on electron-electron interaction.

Our key observation is that the lowest-energy bands of TBG near the magic angle display two distinct features that dominate the transport properties. These are the Dirac nodes at the charge neutrality point and, on the other hand, a set of extended saddle points which, according to experimental\cite{Kim16b,Cao16} and also theoretical\cite{Cea19,Rademaker19} evidence, are close to the Fermi level at half-filling of the two highest (lowest) valence (conduction) bands. The extended saddle points are characterized by a dispersion with very small curvature along the $\Gamma K$ lines that manifests in the straight segments of the Fermi line for the second highest valence band (VB) represented in Fig. \ref{fline}(b). We have argued that this feature could be at the origin of the observed superconductivity of TBG, relying on a Kohn-Luttinger mechanism where the strongly anisotropic screening leads to an attractive interaction between Cooper pairs around the Fermi line.\cite{Gonzalez19}

\begin{figure}[t]
\includegraphics[width=0.49\columnwidth]{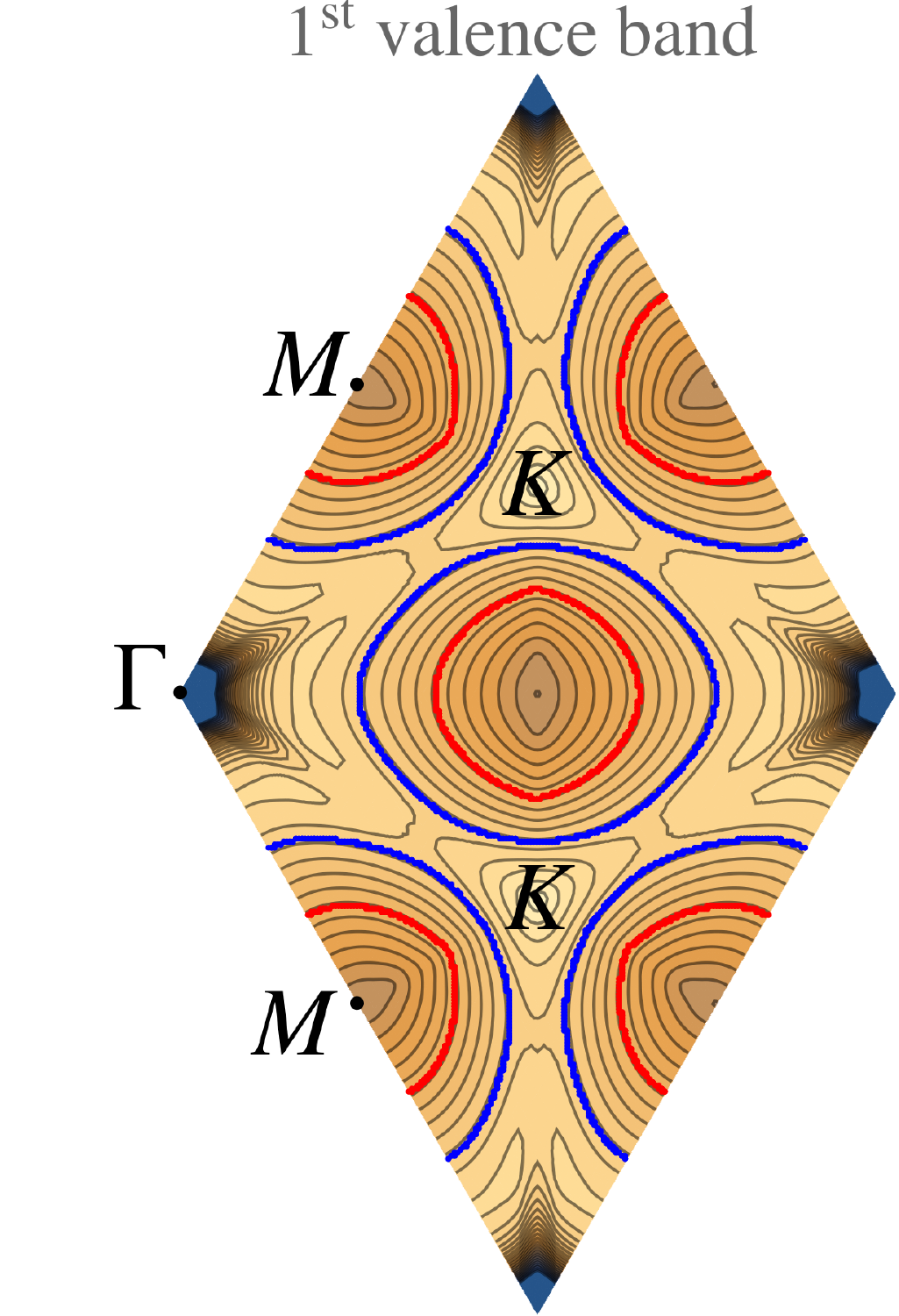}
\includegraphics[width=0.49\columnwidth]{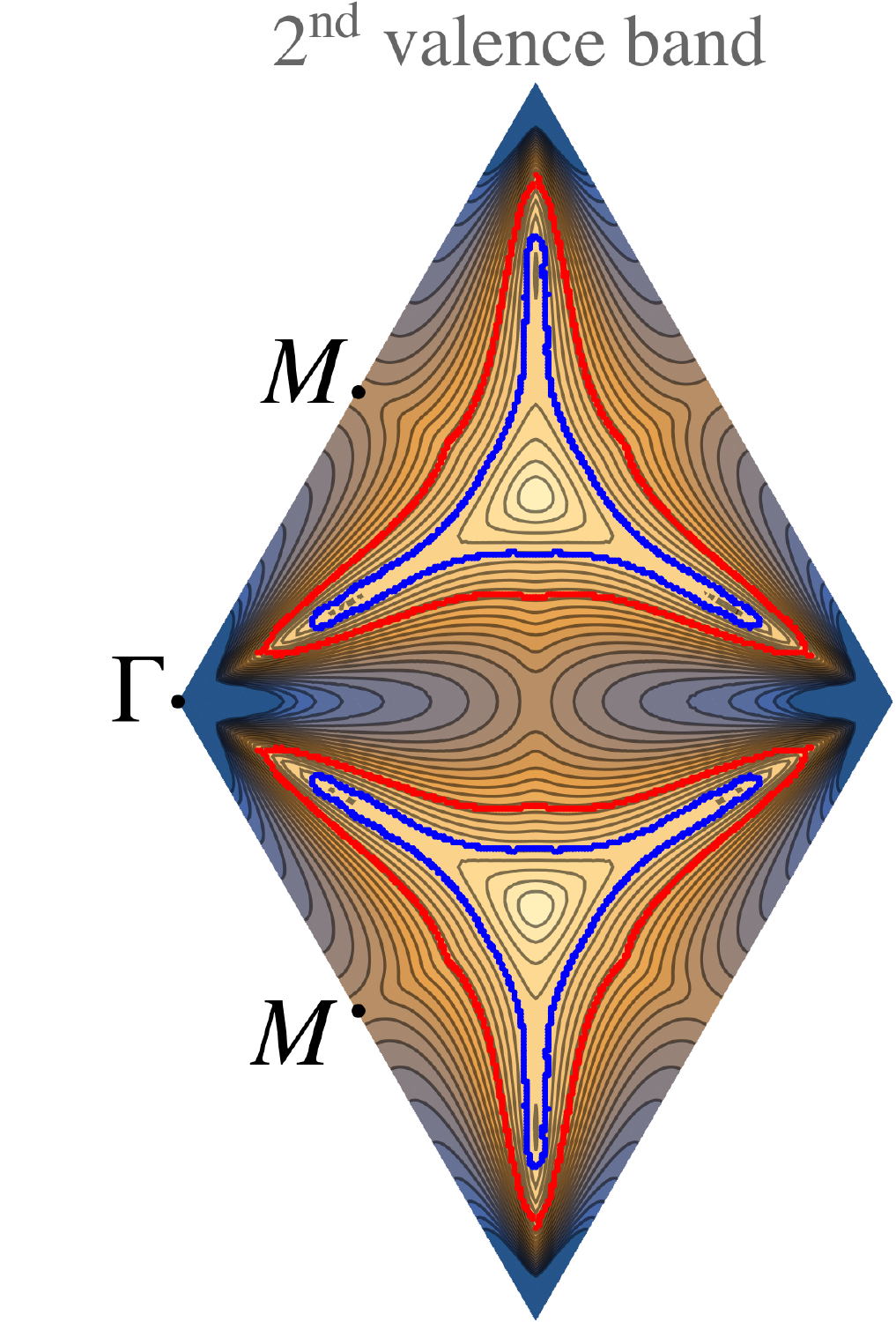}
\\
 \hspace*{1.0cm} (a) \hspace{3.7cm} (b)
\caption{Energy contour maps (with light colors representing higher energies) of the first and the second highest valence band in the Moir\'e Brillouin zone of a twisted graphene bilayer with twist angle $\theta_{28}\approx 1.16^\circ$, showing the Fermi lines for filling levels shifted $-0.2$ meV (blue lines) and $-1.5$ meV (red lines) below the level of the saddle points placed along the $\Gamma K$ lines.  
\label{fline}}
\end{figure}

In this Letter, we show that the decay of the low-energy quasiparticles in the region of flat dispersion (typically within 1 meV around the extended van Hove singularity (vHS)) as well as of those with higher energy (already located in the Dirac cone below the $K$ point) can account for a $T$-linear dependence of the resistivity. The origin of the $T$-linear behavior is, however, different in the two regimes, forcing the appearance of a crossover regime, i.e., a change of slope of the resistivity around a cross-over temperature $T$. This change in the slope is, indeed, observed in the measurements reported in Ref. \onlinecite{Cao19}.

Furthermore, the $T$-linear resistivity observed in the low-temperature regime below $\sim 5$ K must be just one of the many facets revealing the deviation of TBG from the conventional Fermi liquid picture. The reason for such an anomalous behavior lies in the linear growth with energy of particle-hole excitations across the straight segments of the blue Fermi line shown in Fig. \ref{fline}(b). This kind of linear scaling was actually the main assumption in the original proposal of the marginal Fermi liquid paradigm,\cite{Varma89} developed phenomenologically to describe the normal phase of the high-$T_c$ superconductors. Our derivation, therefore, can be seen as a concrete realization of the paradigm, which should manifest itself in other observable features of TBG such as the linear energy scaling of the quasiparticle decay rate or the anomalous temperature dependence of the heat capacity and the thermal conductivity. 

{\it Model.} To model TBG, we will use the commensurable tight-binding model parametrised by the integer $i$ corresponding to the twist angle $\cos\theta_i=\frac{3i^2+3i+0.5}{3i^2+3i+1}$.\cite{Suarez10,Trambly10} The hopping parameters are taken from Ref. \onlinecite{Brihuega12,Moon13} such that the nearest-neighbour intra-layer hopping is set to $t=-2.7$ eV and the vertical interlayer hopping to $t_\perp=0.48$ eV. For a recent review on bilayer systems, see Ref. \onlinecite{Rozhkov16} and also the Supplemental Material (SM).\cite{SI}%\cite{Stauber18b} 

We note that the anomalous transport behavior arises from the peculiar features of the second highest VB of TBG. For small twist angle $\theta \approx 1^\circ$, the first and second highest VBs have coincident vHS at the same energy, but with very different dispersion in the two cases. This dispersion takes the form of a conventional saddle point in the first VB, while that in the second VB has a more distorted shape, with a much smaller curvature along the $\Gamma K$ line than in the orthogonal direction.

In Fig. \ref{fline}, the energy contour maps of the two highest VBs are shown for a bilayer with twist angle $\theta_{28}\approx 1.16^\circ$. Also shown are the Fermi lines for two energies $\Delta \mu$ relative to the vHS. In the second highest VB, they consist of two patches with three lobes each. Notable are the straight segments with quasi one-dimensional dispersion in the second highest VB for $\Delta \mu=-0.2$ meV, which are a reflection of the extended vHS. 

Let us also stress that the energy contour maps of Fig. \ref{fline} with their characteristic Fermi lines around the vHS appear to be rather universal, i.e., independent of the specific details of the underlying microscopic model that may include relaxation effects or slightly different hopping parameters. Remarkably, the same extended vHS are also found in the continuum model,\cite{Lopes07,Mele10,Bistritzer11,Moon12} see Ref. \onlinecite{Gonzalez19}.

\begin{figure}[t]
\includegraphics[width=0.70\columnwidth]{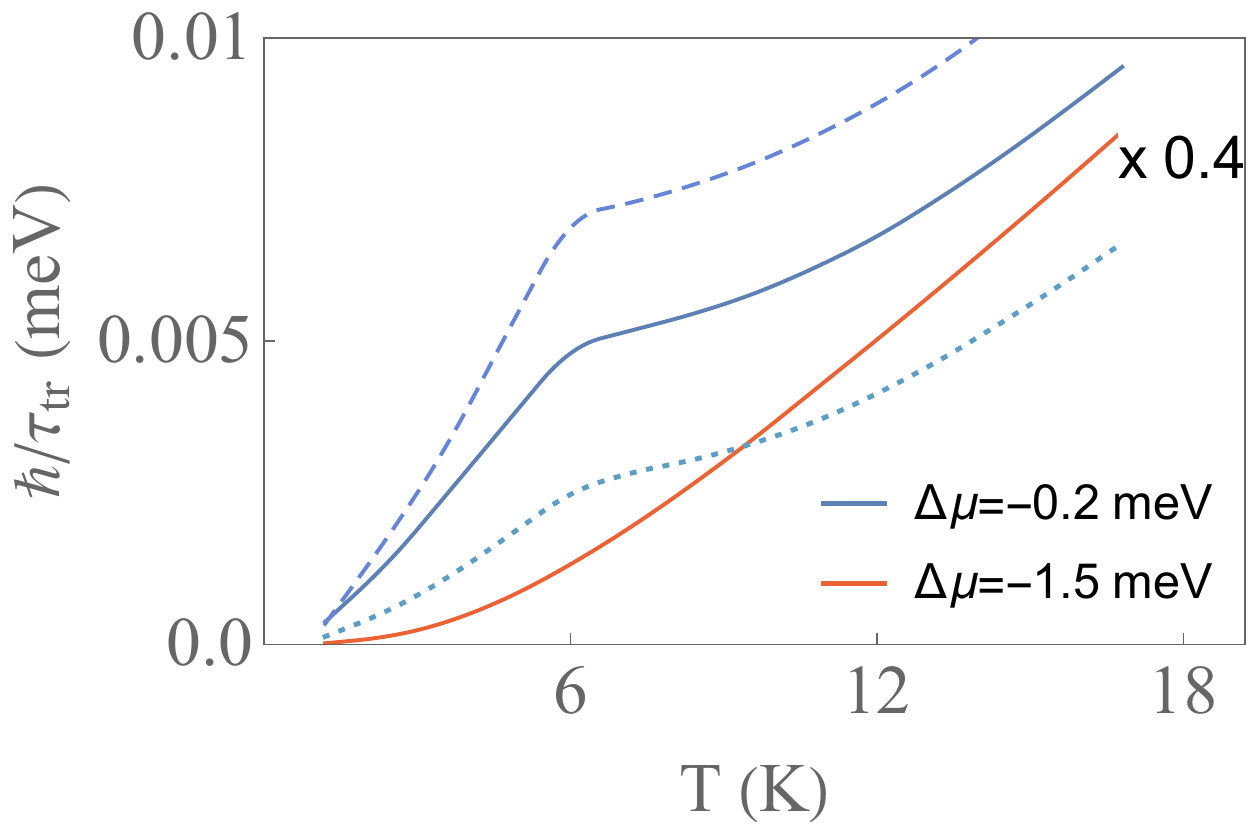}
\caption{Plot of the temperature dependence of the values averaged over the Fermi line (and weighted with the inverse of the square of the Fermi velocity to get dimensions of energy) of $1/\tau_{\rm tr}^{(11)}$ (dashed line), $1/\tau_{\rm tr}^{(22)}$ (solid line) and $1/\tau_{\rm tr}^{(12)}$ (dotted line), when the Fermi level is $0.2$ meV (blue curves) and $1.5$ meV (orange curve, scaled by a factor of 0.4) below the vHS. 
\label{dr02}}
\end{figure}

{\it Resistivity.}
We compute the resistivity relying on the semiclassical formalism of the Boltzmann equation. In this approximation and assuming an on-site Hubbard interaction $U$, the resistivity $\rho_{\bm n}$ in the direction of the unit vector ${\bm n}$ can be expressed as an average over momenta $\k$\cite{Hlubina95}
\begin{align}
\rho_{\bm n} =  &  \rho_0   \frac{ \frac{1}{T} \int \frac{d^2 k}{(2\pi )^2}  \frac{n_F(\varepsilon_\k)}{\tau_{\rm tr} (\k)}  }{  \left(\int \frac{d^2 k}{(2\pi )^2} \frac{\partial n_F(\varepsilon_\k)}{\partial \varepsilon_\k}  ({\bm v}_\k \cdot {\bm n})^2 \right)^2  }    \; ,  
\label{resist}
\end{align}
where $\rho_0 = h/e^2 $ (restoring for a moment Planck's constant), $n_F(\varepsilon_\k)$ is the Fermi-Dirac distribution function at energy $\varepsilon_\k$, the Fermi velocity is ${\bm v}_\k = {\bm \nabla } \varepsilon_\k$, and $1/\tau_{\rm tr}$ stands for the transport decay rate. At this point, it is instructive to discern the different contributions from the decay of quasiparticles in the $i$th VB to the $j$th VB, which lead to partial decay rates
\begin{align}
 & \frac{1}{\tau_{\rm tr}^{(ij)} (\k)} = \: U^2 \int \frac{d^2 k'}{(2\pi )^2} \int_0^{\varepsilon_{\k}^{(i)}} d\omega  \;|\langle i,\k|j,\k'\rangle|^2   \notag \\ 
   &  \times   ( 1-n_F(\varepsilon_{\k'}^{(j)}) ) \; \delta   (\varepsilon_{\k}^{(i)} - \varepsilon_{\k'}^{(j)} - \omega  ) \;  {\rm Im} \chi_{\rm tr}^{(ij)} (\k , \k'; \omega )
\end{align}
with the imaginary part of the transport susceptibility  
\begin{align}
  & {\rm Im} \chi_{\rm tr}^{(ij)} (\k , \k' ; \omega ) =      \label{trs}  \\ 
    &  \int \frac{d^2 p}{(2\pi )^2}  |\langle l,\p|l',\p + \k - \k' \rangle|^2  n_F(\varepsilon_\p^{(l)})  ( 1-n_F(\varepsilon_{\p + \k - \k'}^{(l')})  ) \notag \\
    &    \left[ ({\bm v}_\k^{(i)} + {\bm v}_\p^{(l)} - {\bm v}_{\k'}^{(j)} - {\bm v}_{\p +\k - \k'}^{(l')}) \cdot {\bm n} \right]^2 
        \delta  (\varepsilon_{\p + \k - \k'}^{(l')} - \varepsilon_{\p}^{(l)} - \omega  ) \notag.
\end{align}
Above, we have introduced the eigenvectors $|i,\k\rangle$ corresponding to states in the $i$th highest VB with eigenenergies $\varepsilon_\k^{(i)}$, and the sum over $l,l'$ is implied.\footnote{In practice, we confine the sum in Eq. (\ref{trs}) to intraband processes in the second VB, which is justified as these give rise to the dominant susceptibility arising from electron-hole excitations across the straight segments of the Fermi line in Fig. \ref{fline}(b).} The Hubbard coupling $U$ is the projected on-site interaction onto the second highest VB of the Moir\'e lattice, which we set to $U/2\pi=3$ meV $a_M^2$, $a_M$ being the lattice constant of the superlattice.

The behavior of the resistivity and the transport scattering rate depends on the shift $\Delta \mu $ of the Fermi level with respect to the extended saddle points shown in Fig. \ref{fline}. When the Fermi level is close to the vHS (with a deviation $\Delta \mu$ within $\lesssim 0.5$ meV), the low-energy electron quasiparticles have a dominant decay channel into particle-hole excitations across the straight segments of the blue Fermi line shown in Fig. \ref{fline}(b). These excitations have an approximate linear energy-momentum dispersion, realizing then one of the basic assumptions of the marginal Fermi liquid paradigm. We stress that such a decay channel acts efficiently for the intraband scattering of quasiparticles in both the first and the second highest VB. Consequently, the temperature dependence of $1/\tau_{\rm tr}^{(11)}$ and $1/\tau_{\rm tr}^{(22)}$ for these low-energy quasiparticles (with $\varepsilon_\k$ within $\sim 0.5$ meV around the vHS) is found to be linear in the two bands, as can be seen in Fig. \ref{dr02}. Moreover, interband scattering is shown to provide only a subdominant, softly quadratic correction given by $1/\tau_{\rm tr}^{(12)}$, represented also in Fig. \ref{dr02}.

Beyond a certain temperature, however, the quasiparticles are excited at higher energies away from the vHS, lying on the Dirac cone which is right above the extended saddle points. These quasiparticles can decay into particle-hole excitations not attached to the straight segments of the Fermi line, providing a conventional scattering mechanism which is reflected in the departure from $T$-linear behavior above $T \sim 6$ K in the plot of Fig. \ref{dr02}.

We, therefore, see that the extended saddle-point regime leads to a crossover in the temperature dependence of the transport decay rate, whenever the shift $\Delta \mu$ of the Fermi level is sufficiently small. 
On the other hand, if the Fermi level significantly departs from the vHS (with $\Delta \mu \gtrsim 1$ meV) then the Fermi line recovers a more regular (two-dimensional) shape, as shown by the red lines in Fig. \ref{fline}, and the scattering of electron quasiparticles follows a conventional trend. In this case, the temperature dependence of the transport decay rate has a quadratic behavior starting at low $T$, as shown in Fig. \ref{dr02}.

From the results for the transport decay rate, we obtain the resistivity by applying Eq. (\ref{resist}). At low temperatures, we may assume that only quasiparticles in the energy range of $T$ contribute, so that the resistivity can be computed as an average over the branches of the Fermi line in the two highest VBs. We may discern again different contributions $\rho_{ij}$ from the intraband and interband scattering processes accounted for by $1/\tau_{\rm tr}^{(ij)}$. Decomposing the momentum $\k $ into a longitudinal $k_{\parallel }$ and a transverse $k_{\perp }$ component, we have
\begin{align}
\rho_{ij} \sim   \rho_0  \frac{1}{T} \oint_{{\cal C}_i}  dk_{\parallel } \int \frac{d\varepsilon_{\k}^{(i)}}{v_{\k}^{(i)}} \frac{n_F(\varepsilon_\k^{(i)})}{\tau_{\rm tr}^{(ij)} (\k)}%\;,
\label{res}
\end{align}
${\cal C}_i$ being the Fermi line of the corresponding VB.  

\begin{figure}[t]
\includegraphics[width=0.70\columnwidth]{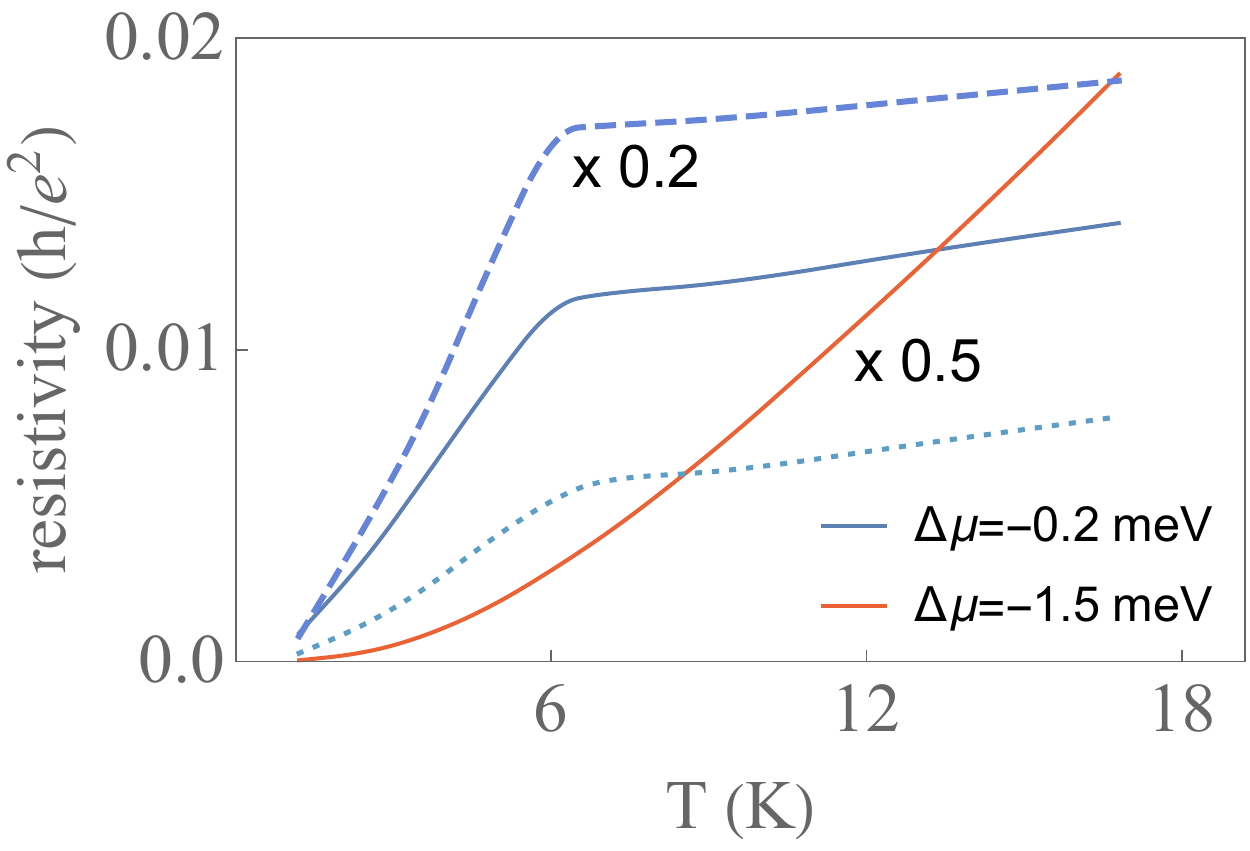}
\caption{Plot of the temperature dependence of $\rho_{11}$ (dashed line, scaled by a factor of 0.2), $\rho_{22}$ (solid line) and $\rho_{12}$ (dotted line), for a shift of the Fermi level $\Delta \mu = -0.2$ meV (blue curves) and $-1.5$ meV (orange curve, scaled by a factor of 0.5) with respect to the level of the vHS. 
\label{resist1}}
\end{figure}

In Fig. \ref{resist1}, one can see that the crossover and the $T$-linear behavior at low temperature of $1/\tau_{\rm tr}^{(11)}$ and $1/\tau_{\rm tr}^{(22)}$ are translated into a similar behavior of the resistivities $\rho_{11}$ and $\rho_{22}$. Interband scattering is found to provide only a subdominant, softly quadratic correction $\rho_{12}$. Above the crossover temperature, however, Eq. (\ref{res}) introduces an additional $T$ dependence, due to the fact that the quasiparticles decaying from the Dirac cone have a reduction of phase space as the energy increases towards the Dirac node at the $K$ point. The $1/T$ factor in Eq. (\ref{res}) is thus not canceled and the temperature dependence of the resistivity becomes approximately linear above the crossover temperature, as shown in Fig. \ref{resist1}.

We note that the magnitude of the resistivity $\rho_{11}$ turns out to be larger than the one originating from the second highest VB. This is a consequence of the fact that the effective bandwidth (for momenta close to the Fermi line) is larger in this latter band. The electronic transport will thus be short-circuited through this second highest VB and it is this band that will dictate the dominant behavior of the resistivity.

As was the case for the transport decay rate, the behavior of the resistivity crucially depends on the shift of the Fermi level with respect to the vHS. This is evidenced in the quadratic behavior of the orange curve in Fig. \ref{resist1} that refers to $\Delta \mu = -1.5$ meV. Nevertheless, for $\Delta \mu = -0.2$ meV, the linear $T$-dependence of the resistivity is quite clear, although with different slope above and below a crossover temperature of the order of $\sim 6$ K. Very suggestively, a change in the slope of the resistivity has also been seen in the experimental observations around half-filling of the Moir\'e unit cell, displaying a larger slope of the linear $T$-dependence below a crossover temperature which is slightly above 5 K in the measurements reported in Ref. \onlinecite{Cao19}. For more details, see the SM.\cite{SI}

{\it Quasiparticle properties.} The linear low-temperature dependence of the transport decay rate can be related to the low-energy behaviour of the electron self-energy $\Sigma (\k, \omega )$. When analyzing this quantity, it is convenient to discern the different contributions to its imaginary part from the decay of quasiparticles in the $i$th VB to the $j$th VB. These can be computed in terms of the conventional electron-hole susceptibility $\chi (\q, \omega ) $ as
\begin{align}
  &  {\rm Im} \: \Sigma^{(ij)} (\k, \omega ) =  -U^2 \int \frac{d^2 p}{(2\pi)^2} \int_{-\infty }^{\infty } d\omega_p \: |\langle i,\k|j,\p \rangle|^2  \notag  \\ 
      &   \times  {\rm sgn}(\omega_p ) \: \delta(\omega_p - \varepsilon_{\p}^{(j)}) \; {\rm Im} \: \chi (\k - \p, \omega - \omega_p)\;.
\label{ims}
\end{align}
The real part of the self-energy can be obtained by making use of the Kramers-Kronig relation, which in this case takes the form
\begin{align}
{\rm Re} \: \Sigma (\k, \omega ) = \frac{2\omega }{\pi } \int_0^\infty d\Omega \: \frac{{\rm Im} \: \Sigma (\k, \Omega )}{\Omega^2 - \omega^2}\;.
\end{align}

\begin{figure}[t]
\includegraphics[width=0.49\columnwidth]{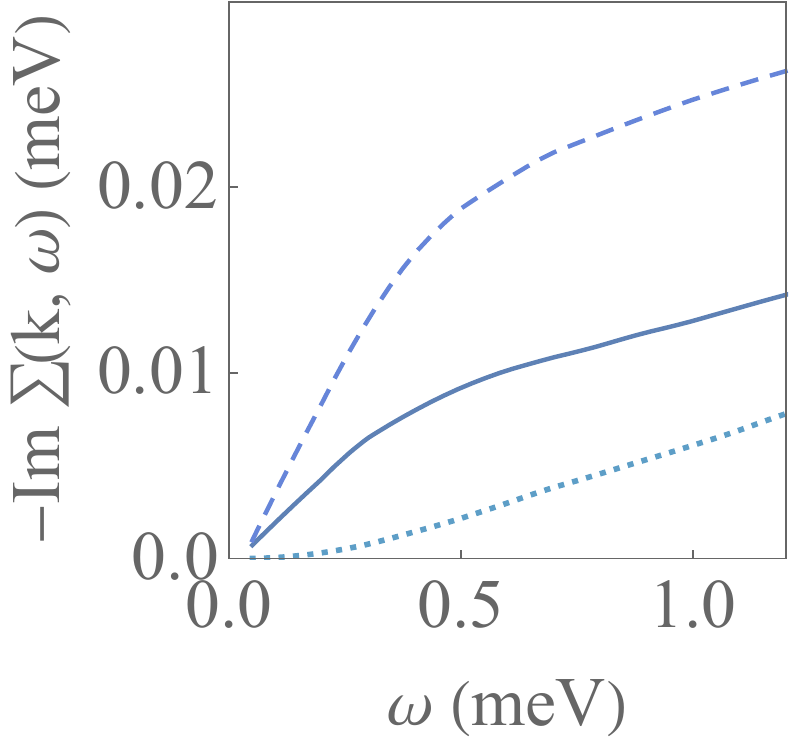}
\includegraphics[width=0.47\columnwidth]{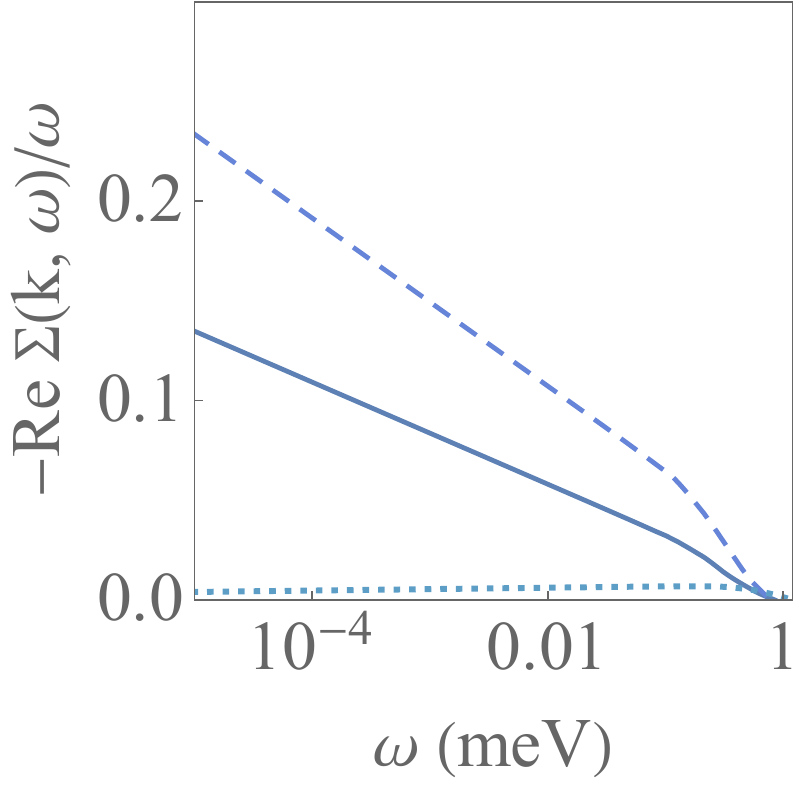}
\\
 \hspace*{1.0cm} (a) \hspace{3.7cm} (b)
\caption{Plot of the frequency dependence of the real and the imaginary part of the values averaged over the Fermi line of $\Sigma^{(11)}$ (dashed line), $\Sigma^{(22)}$ (solid line) and $\Sigma^{(12)}$ (dotted line), for a shift of the Fermi level $\Delta \mu = -0.2$ meV below the vHS.
\label{realself}}
\end{figure}

When the Fermi level is close to the vHS, the imaginary part of $\Sigma^{(11)}$ and $\Sigma^{(22)}$ computed from Eq. (\ref{ims}) has a linear dependence on the frequency $\omega $ which is similar to the low-temperature dependence of $1/\tau_{\rm tr}^{(11)}$ and $1/\tau_{\rm tr}^{(22)}$, as can be seen in Fig. \ref{realself}(a). As shown in the same figure, the effect of interband scattering is reflected in a small correction from $\Sigma^{(12)}$ at low frequencies. Consequently, the dependence of the real part of the self-energy on $\omega $ displays a significant deviation from linear behavior, with a logarithmic correction which is another evidence of the departure from Fermi liquid behavior. This is shown in Fig. \ref{realself}(b), which represents the average over the Fermi line of the real part of the different contributions $\Sigma^{(ij)} (\k, \omega )/\omega $ when the Fermi level is shifted by $\Delta \mu = -0.2$ meV with respect to the vHS.

We observe that the real part of the dominant intraband contributions to the self-energy behaves as ${\rm } \: \Sigma (\k, \omega ) \sim \omega \: \log(\omega )$ at low frequencies, which amounts to state that the electron quasiparticles are progressively attenuated when approaching the Fermi level. The dressed electron propagator becomes
\begin{align}
G (\k, \omega ) = \frac{1}{\omega - \varepsilon_\k - \Sigma (\k, \omega )} \sim  \frac{\Delta }{\omega - \varepsilon_\k + i\gamma \omega }
\end{align}
after rewriting the self-energy corrections in terms of the quasiparticle weight $\Delta $ and the imaginary shift $i\gamma \omega $ of the quasiparticle pole. The quasiparticle weight is suppressed following the low-energy scaling $\Delta \sim 1/|\log(\omega )|$, which is the hallmark of the marginal Fermi liquid behavior.\cite{Varma89,Littlewood91} For more details, see the SM.\cite{SI}

{\it Heat capacity and thermal conductivity.}
The anomalous behavior of the electron quasiparticles has also a significant impact on the temperature dependence of observables like the heat capacity. This is obtained from the entropy $S$, which can be expressed as an integral along the Fermi line by decomposing again the momentum $\k $ into longitudinal $k_{\parallel }$ and transverse $k_{\perp }$ components\cite{SI,Abrikosov63}:
\begin{align}
\frac{S}{A}  \approx \frac{1}{2 \pi^2} \frac{1}{T} \oint  \frac{dk_{\parallel }}{v_{\k}}  \int_{-\infty }^{\infty } d\omega \: \omega \frac{\partial n_F(\omega)}{\partial \omega }  \left( \omega - {\rm Re} \: \Sigma (\k, \omega ) \right)  \label{entr},
\end{align}
$A$ being the area of the system. Then, by absorbing the temperature $T$ into a dimensionless variable $\omega/T$ in the integrand of Eq. (\ref{entr}), we see that the anomalous scaling of the electron self-energy translates into the dominant scaling behavior $S \sim T \: |\log(T)|$.

The heat capacity $C$ is obtained by taking the derivative of $S$ with respect to $T$ and it inherits, therefore, the logarithmic correction that we find in the entropy:
\begin{align}
C =  T \frac{\partial }{\partial T} \frac{S}{A}  \sim  T \: |\log(T)|
\end{align}
We see therefore that the logarithmic correction to the heat capacity holds in the same range of anomalous behavior of the self-energy plotted in Fig. \ref{realself}, which corresponds in temperature to the range $T \lesssim 10$ K.

The logarithmic correction of the heat capacity has also a direct translation into the temperature dependence of the thermal conductivity $\kappa $. This quantity is related to the heat capacity through the thermal diffusivity $\alpha $ according to the formula $\kappa = \alpha C $. The thermal diffusivity is in turn proportional to the mean free path of the energy carriers.\cite{Ziman72} When the Fermi level is close to the vHS, we can apply the linear low-temperature dependence we have found in the transport scattering rate to estimate the mean free path.\cite{Varma89} This implies that
\begin{align}
\kappa (T) = \alpha  C \sim |\log (T)|\;. 
\label{kap}
\end{align}
This anomalous scaling should be observable down to the temperature scale at which the transport starts to be dominated by the scattering from disorder (impurities or lattice defects) in the twisted bilayer. Above that scale, the ratio between the thermal conductivity and the electrical conductivity should be also affected by the logarithmic correction from Eq. (\ref{kap}), thus leading to a modification of the Wiedemann-Franz law.\cite{Varma89}

{\it Long-range interaction.} So far, we have considered the case of a strongly screened Hubbard interaction that can be interpreted as some effective parameter $U$ that also includes the dielectric constant of the substrate. Within the continuum model,\cite{Lopes07} we have further investigated the relaxation time for long-range interaction including screening effects\cite{Giuliani82} that come from the top and back gate as well as from internal self-screening. Interestingly, apart from the linear vs quadratic behavior as function of the chemical potential relative to the vHS, we obtain relaxation times comparable to the Planckian limit for gate distances $D=15$ nm, see Fig. 4 of the SM.\cite{SI} Within the same framework, we have discussed the influence of the relaxation time to the quasi-localised plasmonic modes\cite{Stauber16}, which also leads to a $T$-linear behavior proportional to the density of states, see SM.\cite{SI} 

{\it Summary.} Relying on a tight-binding model, we have been able to obtain a linear temperature dependence of the resistivity for filling factors around the vHS in the two highest VBs of TBG, in the framework of a model with on-site Hubbard interaction $U$. At low temperatures, the linear behavior of the resistivity can be traced back to the more general frequency dependence of the electron Green's function, characterized by a logarithmic correction indicating marginal Fermi liquid behavior. We thus predict that fingerprints of a marginal Fermi liquid should also be present in the heat capacity and the thermal conductivity. Observing these features experimentally may be a way to discriminate between the electron-electron and the electron-phonon interaction as the possible driving force for the superconducting state as well as for the unconventional normal state found near half-filling of the two highest VBs in TBG.

Finally, we stress that the scaling laws we have discussed persist when the Coulomb interaction is extended to get a finite spatial range. In that case, quantitative predictions about the different observables may be greatly enhanced and, specially in the limit of a long-range Coulomb interaction (with appropriate internal screening), a regime of nearly Planckian resistivity can be reached, with the transport decay rate approaching the bound given by $T/\hbar $.

{\it Acknowledgements.}
This work has been supported by Spain's MINECO under Grant No. FIS2017-82260-P as well as by the CSIC Research Platform on Quantum Technologies PTI-001. 

\begin{widetext}
{\bf\huge Supplemental Material}\\
\section{Tight-binding Hamiltonians}
For the calculation of the relaxation time and resistivity, we rely on the use of a tight-binding model. We adopt a general formulation of the tight-binding approach with Hamiltonian
\begin{align}
H = -\sum_{\langle i,j\rangle} t_{\parallel} (\r_i-\r_j) \; (a_{1,i}^{\dagger}a_{1,j}+h.c.) - \sum_{\langle i,j\rangle} t_{\parallel} (\r_i-\r_j) \; (a_{2,i}^{\dagger}a_{2,j}+h.c.) - \sum_{(i,j)} t_{\perp}(\r_i-\r_j) \; (a_{1,i}^{\dagger}a_{2,j}+h.c.)\;. 
\label{tbh}
\end{align}
The sum over the brackets $\langle...\rangle$ runs over pairs of atoms in the same layer (1 or 2), whereas the sum over the curved brackets $(...)$ runs over pairs with atoms belonging to different layers. $t_{\parallel} (\r)$ and $t_{\perp} (\r)$ are hopping matrix elements which have an exponential decay with the distance $|\r|$ between carbon atoms. A common parametrization is based on the Slater-Koster formula for the transfer integral\cite{Moon13} 
\begin{align}
-t(\d)=V_{pp\pi}(d)\left[1-\left(\frac{\d\cdot\e_z}{d}\right)^2\right]+V_{pp\sigma}(d)\left(\frac{\d\cdot\e_z}{d}\right)^2
\end{align}
with
\begin{align}
V_{pp\pi}(d)=V_{pp\pi}^0\exp\left(-\frac{d-a_0}{r_0}\right)\;,
V_{pp\sigma}(d)=V_{pp\sigma}^0\exp\left(-\frac{d-d_0}{r_0}\right)\;,
\end{align}
where $\d $ is the vector connecting the two sites, $\e_z$ is the unit vector in the $z$-direction, $a_0 $ is the C-C distance and $d_0$ is the distance between layers. A typical choice of parameters is given by $V_{pp\pi}^0=-2.7$ eV, $V_{pp\sigma}^0=0.48$ eV and $r_0=0.319 a_0$ \cite{Moon13}. In particular, we have taken these values to carry out the analysis reported in the main text. For an alternative comparison between the continuous and the tight-binding model, see Ref. \cite{Stauber18b}.

\begin{figure}[h]
\includegraphics[width=0.20\columnwidth]{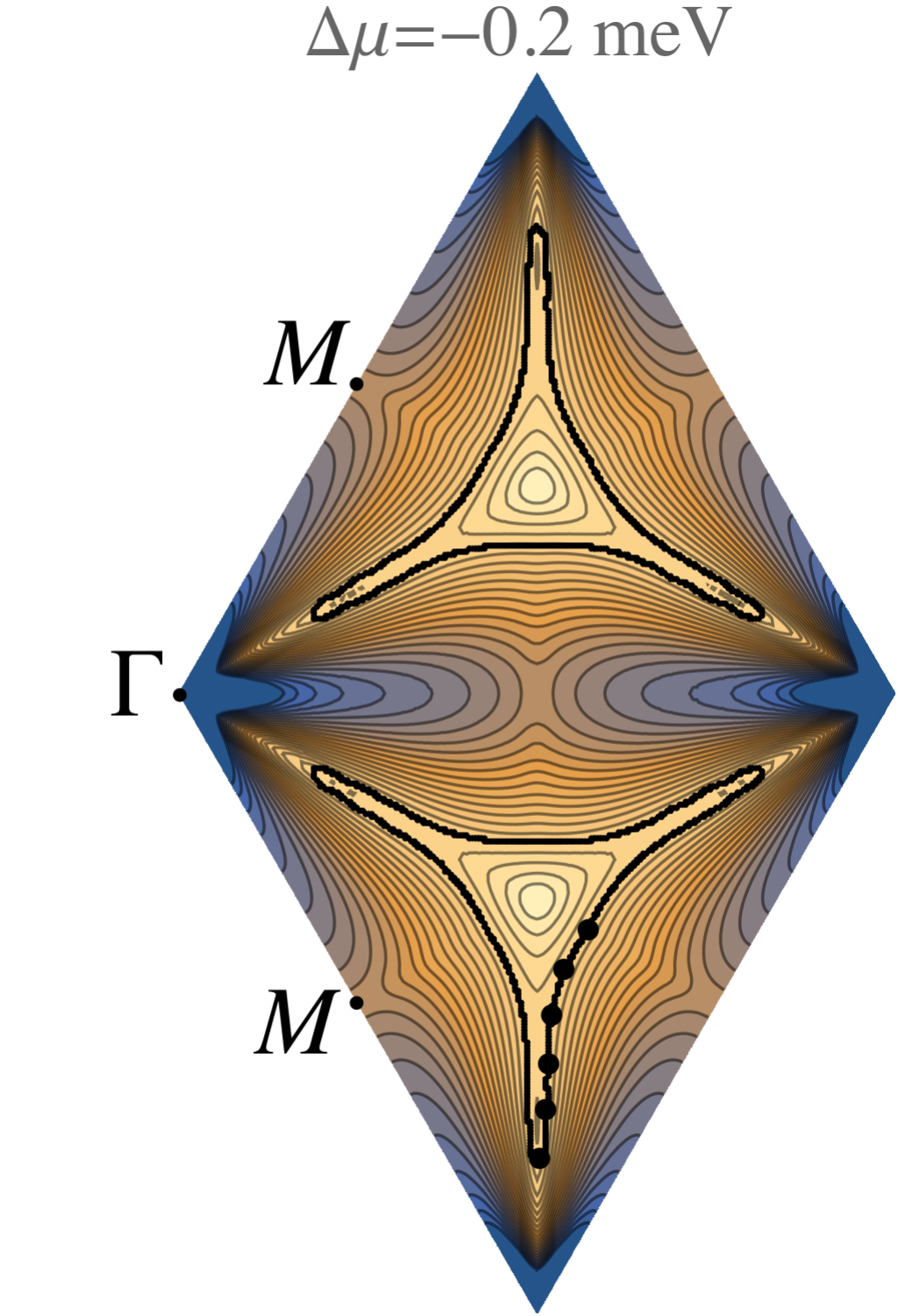}
\includegraphics[width=0.20\columnwidth]{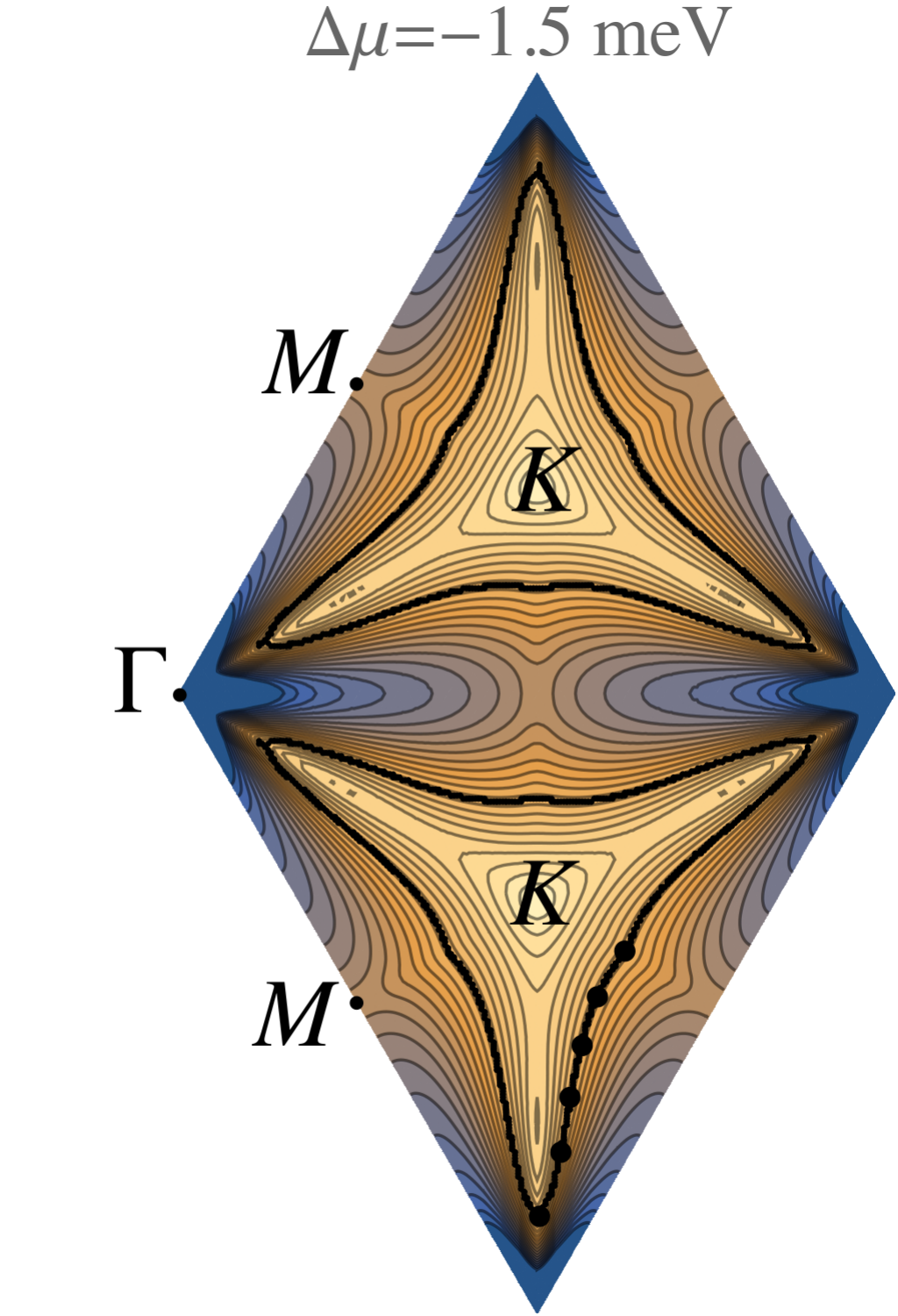}
\caption{Energy contour maps of the second highest valence band in the Moir\'e Brillouin zone of a twisted graphene bilayer with twist angle $\theta_{28}\approx 1.16^\circ$, showing the Fermi lines for filling levels shifted $-0.2$ meV (left) and $-1.5$ meV (right) below the level of the saddle points placed along the $\Gamma K$ lines.  
\label{fline}}
\end{figure}

\section{Transport decay rate of quasi-particles at the Fermi line} 
At low temperature, the transport decay rate is dominated by electron quasiparticles close to the Fermi line. In Fig. \ref{fline}, we show the Fermi line for two different chemical potentials $\Delta \mu$ taken with respect to the level of the van Hove singularity (vHS) arising from the saddle points at the $\Gamma K$ line. Also indicated are the discrete points on the Fermi line for which explicit calculations are here illustrated.

In Fig. \ref{dr02}, we show the transport decay rate as function of temperature, computed according to the expression (2) in the main text, for the different points on the Fermi line indicated in Fig. \ref{fline}. As can be appreciated, the behavior depends crucially on the value of $\Delta \mu $, turning from linear to quadratic at low temperatures as the Fermi level deviates significantly from the vHS.

\begin{figure}[h]
\includegraphics[width=0.40\columnwidth]{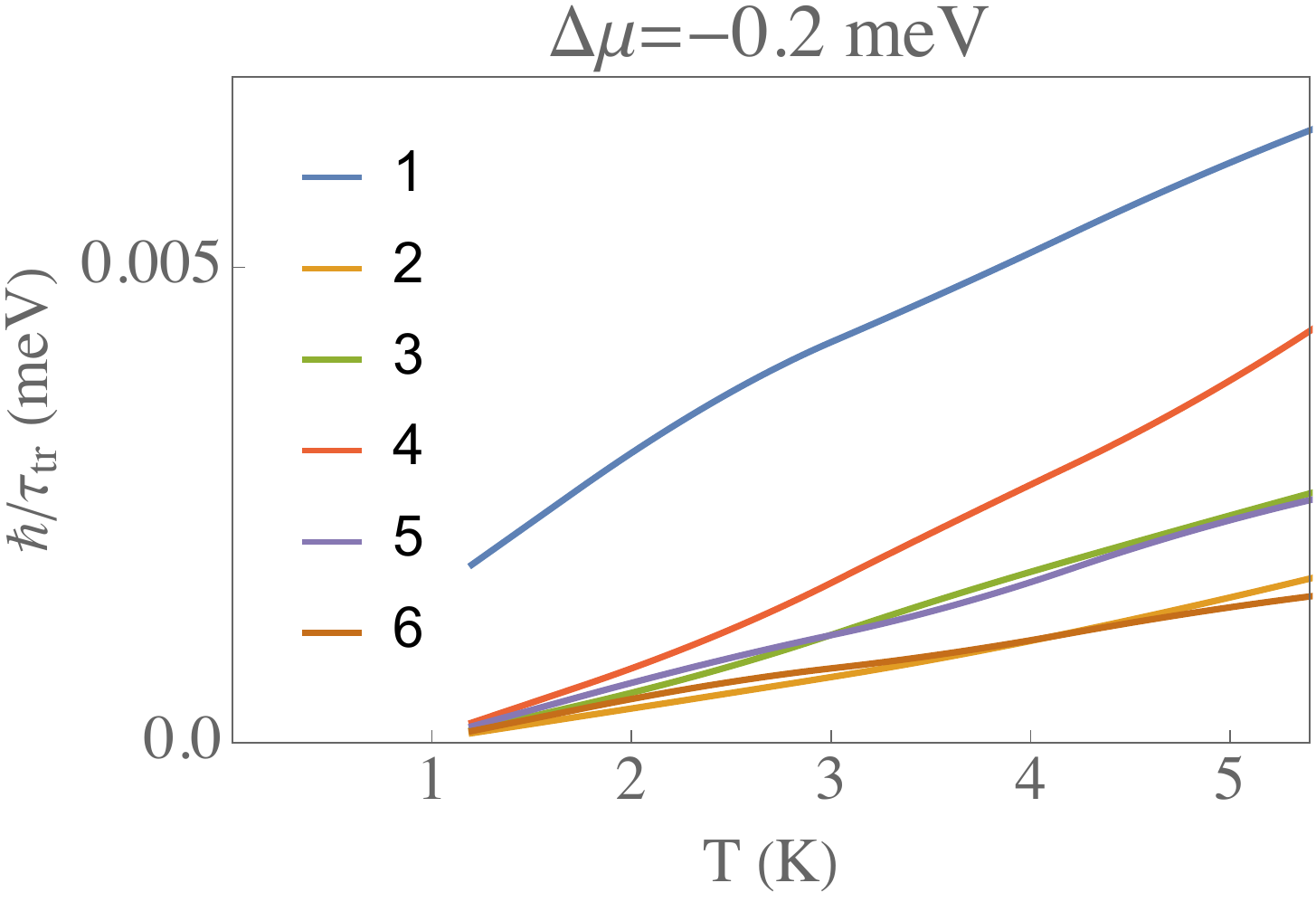}
\hspace{1cm}
\includegraphics[width=0.40\columnwidth]{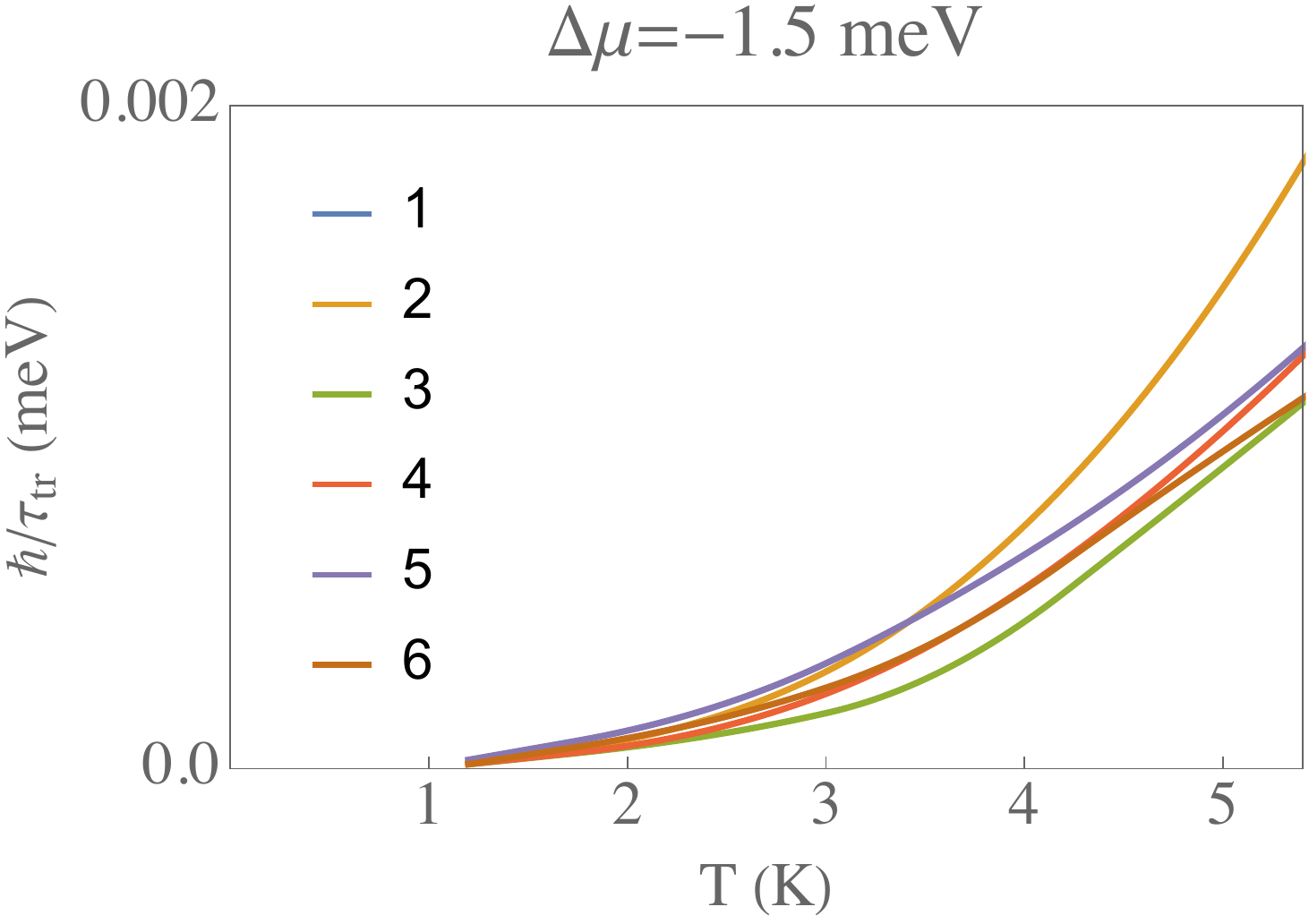}
\\
 \hspace*{1.0cm} (a) \hspace{7.7cm} (b)
\caption{Plot of the temperature dependence of $1/\tau_{\rm tr}$ (weighted with the inverse of the square of the Fermi velocity to get dimensions of energy) when the Fermi level is $0.2$ meV (left) and $1.5$ meV (right) below the vHS, for six points along de Fermi line following the sequence shown in Fig. \ref{fline}, from the farthest position (1) to the closest location to the $K$ point (6). The on-site Hubbard interaction is taken as $U/(2\pi ) = 3$ meV $a_M^2$, $a_M$ being the lattice constant of the superlattice.    
\label{dr02}}
\end{figure}

\section{Quasi-particle properties at the Fermi line} 
Also for the self-energy, we can analyse the low-energy behaviour for different quasi-particles on the Fermi line. This is seen in Fig. \ref{imself}(a), which represents the imaginary part of the self-energy $\Sigma (\k, \omega )$ as function of $\omega $, computed according to the expression (5) in the main text, for the points on the Fermi line indicated in Fig. \ref{fline}. The linear behavior at low frequencies is consistent with the low-temperature dependence of the transport decay rate shown in Fig. \ref{dr02} for $\Delta \mu = -0.2$ meV.
\begin{figure}[h]
\includegraphics[width=0.43\columnwidth]{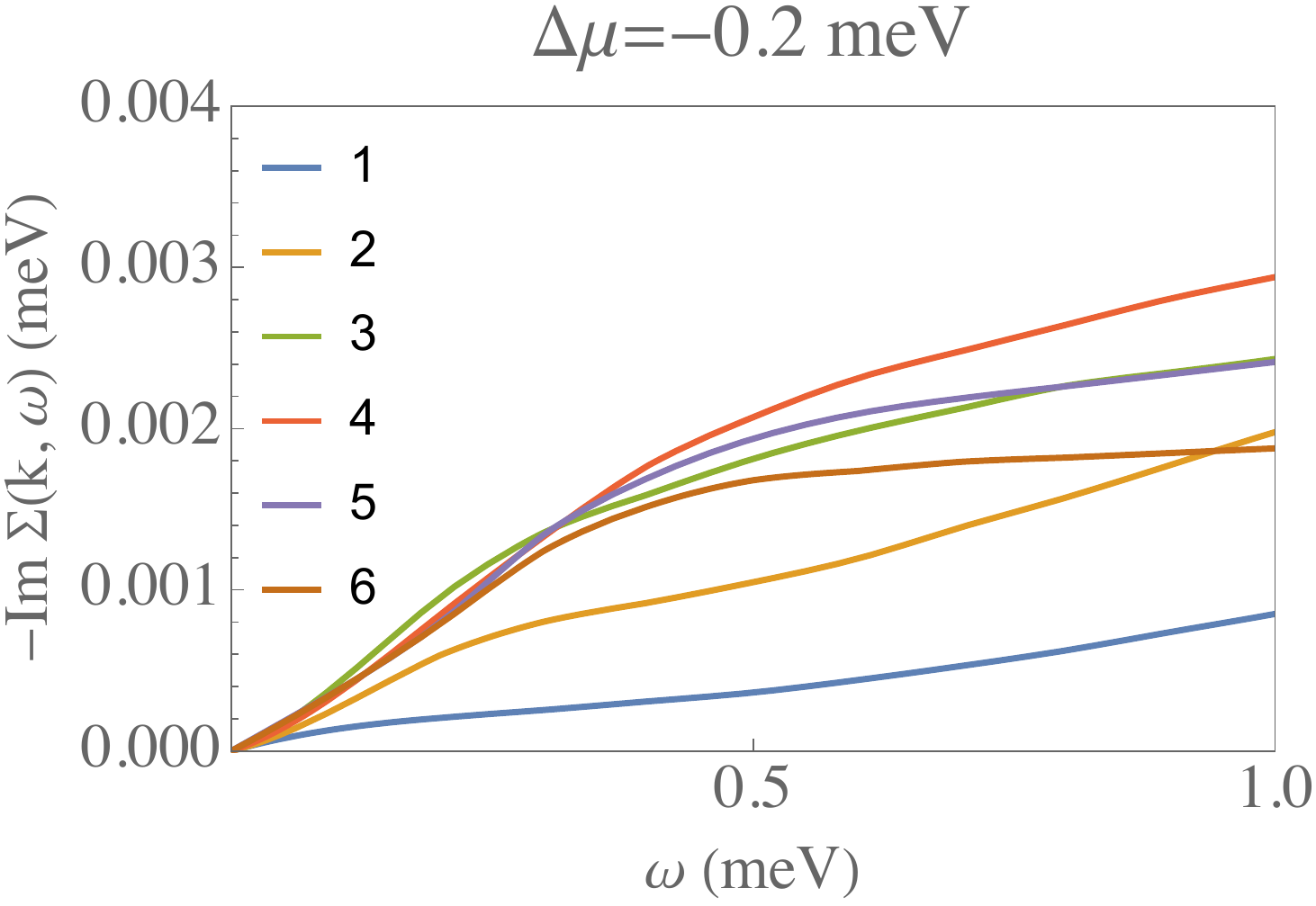}
\hspace{1cm}
\includegraphics[width=0.40\columnwidth]{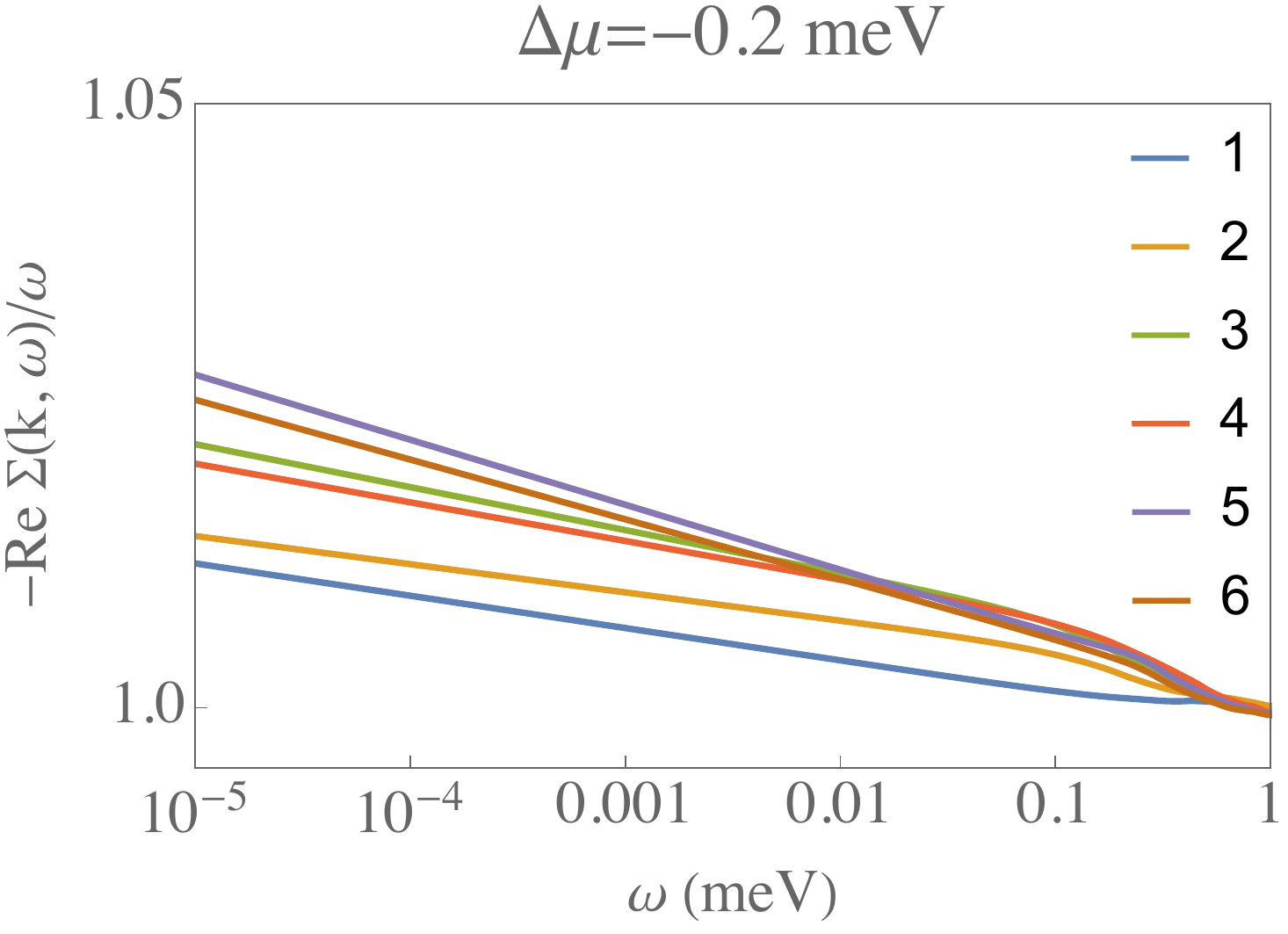}
\\
 \hspace*{1.0cm} (a) \hspace{7.7cm} (b)
\caption{Plot of the frequency dependence of the imaginary (left hand side) and real (right hand side) part of $\Sigma (\k, \omega )$ for six points along de Fermi line following the sequence shown in Fig. \ref{fline}, from the farthest position (1) to the closest location to the $K$ point (6). The Fermi level is $0.2$ meV below the vHS and the on-site Hubbard interaction is taken as $U/(2\pi ) = 3$ meV $a_M^2$, $a_M$ being the lattice constant of the superlattice. 
\label{imself}}
\end{figure}

From the imaginary part of $\Sigma (\k, \omega )$, we can compute the real part of the self-energy by applying the Kramers-Kronig relation in Eq. (6) of the main text. The results corresponding to the different curves in Fig. \ref{imself}(a) are represented in Fig. \ref{imself}(b), which shows a clear logarithmic correction consistent with the linear dependence at low frequencies of the respective imaginary counterparts.

\section{Umklapp processes}
Umklapp processes define scattering events for which the final momentum lies outside the first Brillouin zone. The final momentum and its corresponding energy can be mapped back onto the first Brillouin zone, but this is not allowed for its wave function. Nevertheless, mapping also the eigenvectors onto the first Brillouin zone facilitates the numerical calculations and this approximation leads to a susceptibility that is periodic on the first Brillouin zone. 

The above approximation has been employed in the calculations of the main text and shall here be discussed for the continuum model, i.e., we compare it to the exact result. Another approximation would simply neglect all scattering processes that lie outside the first Brillouin zone. In Fig. \ref{Susceptibility}, we see that both approximations coincide in the case of the susceptibility for small wave numbers. All protocols are thus consistent with our main assumption, i.e., the marginal Fermi liquid behavior is caused by small momentum transfer along the quasi-one dimensional segments of the Fermi line. The same is true for the corresponding relaxation times.

\begin{figure}[t]
\includegraphics[width=0.99\columnwidth]{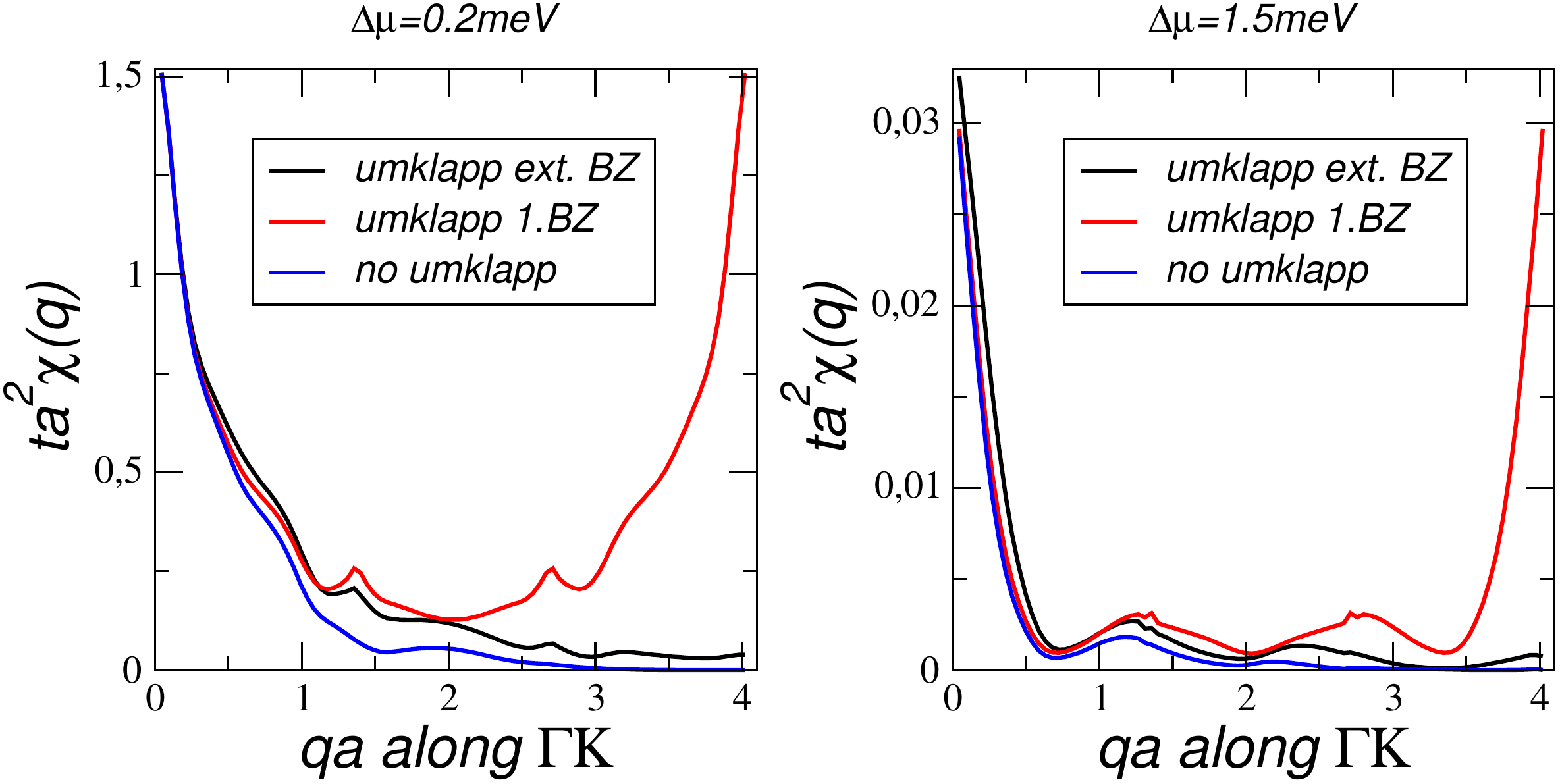}
\caption{Full susceptibility of the second highest VB (black line) for two different chemical potentials relative to the vHS $\Delta\mu=0.2$meV (left) and $\Delta\mu=1.5$meV (right). Also shown the susceptibilities that include approximate treatments of umklapp processes.
\label{Susceptibility}}
\end{figure}

\section{Relaxation time for effective dielectric media}
In the main text, we have assumed a strongly screened Hubbard interaction $U$, valid for gates close to the twisted bilayer sample. Here, we will outline the formalism including screening effects within the $G_0W$-approximation. We will first discuss the case of a dielectric function due to long-ranged Coulomb interaction and then also estimate the effect of localised plasmonic modes predicted in TBG.\cite{Stauber16}

\subsection{Relaxation time for long-ranged interaction.}
For gates further away, we expect also effects from the long-ranged Coulomb potential to become important. In this case, we calculate the scattering rate by incorporating the intrinsic screening effects within the $G_0W$-approximation of the self-energy, starting from the Coulomb potential, screened by a bottom and top gate at distance $D$:\cite{Cea19}
\begin{align}
 v_q=\frac{e^2}{2\epsilon_0\epsilon q}\frac{1-e^{-qD}}{1+e^{-qD}}\;.
\end{align} 
We will set $\epsilon=5$, the approximate value for hBN. 

The relaxation time within the $G_0W$-approximation at finite temperature for a quasiparticle (hole) state with $\Delta=E_\p\mu$ is given by\cite{Giuliani82}   
\begin{align}
\frac{1}{\tau(\Delta)}&=\int_{-\infty}^{\infty}\frac{d\omega}{2\pi}f(\omega)\frac{1}{A}\sum_\q v_q|\langle \p|\p+\q\rangle|^2\Im\left(\frac{1}{\epsilon(\q,\omega)}\right)\delta\left(\hbar\omega-(E_\p-E_{\p+\q})\right)\notag
\end{align}
which involves the dielectric function within the RPA
\begin{align}
\epsilon(\q,\omega)=1-v_q\chi(\q,\omega)
\end{align}
with the polarisability $(g_s=g_v=2)$
\begin{align}
\chi(\q,\omega)=\frac{g_sg_v}{A}\sum_\k|\langle \k|\k+\q\rangle|^2\frac{n_F(E_{\k})-n_F(E_{\k+\q})}{\hbar\omega-(E_{\k+\q}-E_{\k})+i0}\;,
\end{align}
and the temperature-dependent weight factor
\begin{align}
\label{f}
f(\omega)=\frac{\coth(\beta\hbar\omega/2)-\tanh(\beta(\hbar\omega-\Delta)/2)}{1+e^{-\beta\Delta}}\;.
\end{align}
We further defined the eigenstates $|\p\rangle$, the Fermi function $n_F(E)=(e^{\beta E}+1)^{-1}$, the inverse temperature $\beta=1/(k_BT)$. 

The $G_0W$-approximation requires the knowledge of the real and imaginary parts of the susceptibility and, in order to speed up the calculations, we have worked with the less time-consuming continuum model.\cite{Lopes07} In Fig. \ref{ScatteringRateLongRange}, we show the resulting scattering rate for  different chemical potentials relative to the vHS as function of the temperature. Interestingly, for larger gate distances $D\sim15$nm the results are close to the Planckian scattering rate $\hbar/\tau=0.086meV\cdot T[K]$ indicated as dashed line which is in good agreement to the experimental findings of Ref. \onlinecite{Cao19}.

\begin{figure}[t]
\includegraphics[width=0.99\columnwidth]{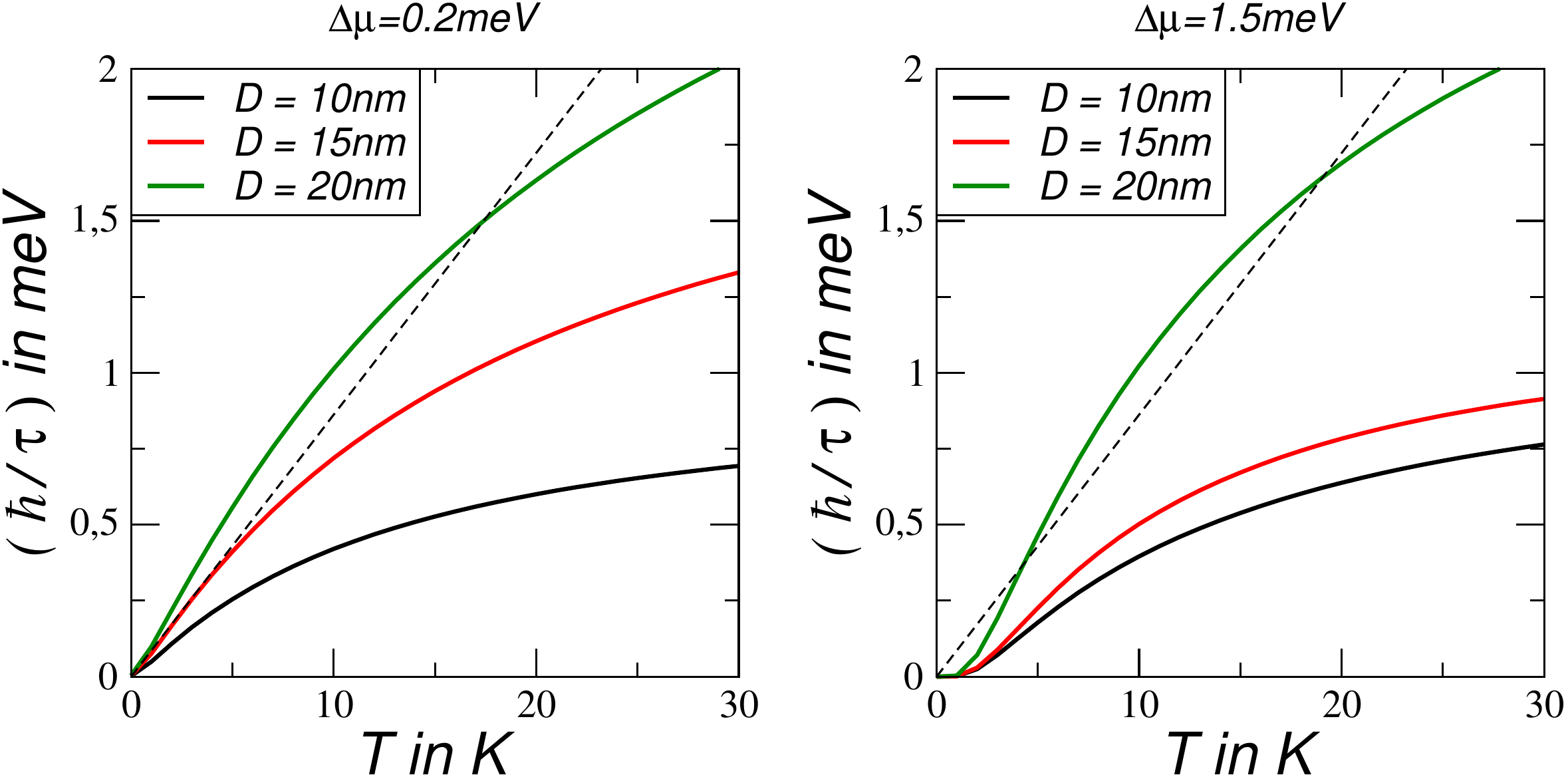}
\caption{The scattering rate $\hbar/\tau$ of TBG with $i=29$ as function of the temperature for two chemical potentials around the vHS and screened long-ranged interaction with surrounding dielectric material $\epsilon=5$. $D$ denotes the distance of TBG to the top and bottom gate. The dashed line indicates the Planckian scattering rate $\hbar/\tau=0.086meV\cdot T[K]$.
\label{ScatteringRateLongRange}}
\end{figure}

What is seen independent of the gate distance $D$ is that for a chemical potential close to the van Hove singularity there is a linear low-temperature behaviour ($\Delta\mu\llless0.2$) in contrary to the quadratic low-temperature behaviour for $\Delta\mu\gsim1.5$meV. Also seen for all curves is the crossover to a different quasi-linear temperature regime for $T_{cr}\sim5$K.

\subsection{Relaxation time from collective modes}
For temperatures larger than the band-gap, the system is expected to reach the classical regime, characterised by a linear behaviour of the resistivity and dominated by the thermal charge fluctuations. In Fig. \ref{LossFunction}, we show the loss function $S(\omega)=$-Im$\epsilon^{-1}(\q,\omega)$ for $|\q|a=0.02$ in the $KK'$-direction for $\epsilon_0=4.8$ for various twist angles. The peak resembles a true plasmonic resonance as discussed in Ref. \cite{Stauber16} which shifts to smaller energies with decreasing twist angles. 

For $i=29$, also a Lorentzian fit is shown with 
\begin{align}
\tilde S(\omega)=\frac{2C\gamma}{(\omega-\omega_0)^2+\gamma^2}\to2\pi C\delta(\omega-\omega_0)\;,
\end{align}
where $\hbar C=3$meV, $\hbar\gamma=4$meV, and $\hbar\omega_0\sim8$meV. This yields the following scattering rate for $U=5meVa_M^2$ and $\mu$ close to the van Hove singularity, i.e., $U\rho(\mu)=4$:
\begin{align}
\frac{1}{\tau}=0.16\frac{k_BT}{\hbar}
\end{align}

With the plasmon energy $\hbar\omega_0/t\approx0.003$, the crossover temperature corresponds to $100$K which is clearly too high to explain the experiments of \cite{Cao19}. But this limit is imposed by the accuracy of our numerical solution and we expect a linear behaviour for the resonance as indicated by the red line in the inset. In fact, the plasmonic resonance is related to the band-width of the lowest valence/conduction bands which is around 1meV resp. 10K. 

The quasi-particle relaxation time will be mainly determined by the specific form of the loss function which was discussed in Ref. \cite{Stauber16} for small angle twisted bilayer graphene. There, a quasi-flat mode was found for small twist angles, independent of moderate doping-levels related to the localised states around the $AA$-region. As a first approach, we can thus approximate the loss function by the following analytical function:
\begin{align}
\Im\left(\frac{1}{\epsilon(\q,\omega)}\right)=2\pi C\delta(\omega-\omega_0)
\end{align}
with some suitable constant $C$ which permits for an analytical solution of the relaxation time neglecting the wave function overlap $|\langle \p|\p+\q\rangle|^2$. We get
\begin{align}
\frac{1}{\tau(\Delta)}=UCf(\omega_0)\rho(E_\p-\hbar\omega_0)
\end{align}
where $\rho(\omega)$ denotes the density of states. For large temperature, $\omega_0\ll k_BT$

We expect that the decay rate is dominated by the electron-plasmon coupling active for quasi-particle energies $\Delta\approx\omega_0$. Then, for sufficiently large temperatures $k_BT\gg\hbar\omega_0$, we obtain
\begin{align}
\frac{1}{\tau(\omega_0)}&\approx UC\rho(\mu)\frac{k_BT}{\hbar\omega_0}
\end{align}
Our approach thus yields the observed linear $T$-resistivity above some energy scale $\hbar\omega$. Furthermore, the prefactor is governed by the density of states which is decreasing as one approaches  the regime of half-filling from below. 

\begin{figure}[t]
\includegraphics[width=0.99\columnwidth]{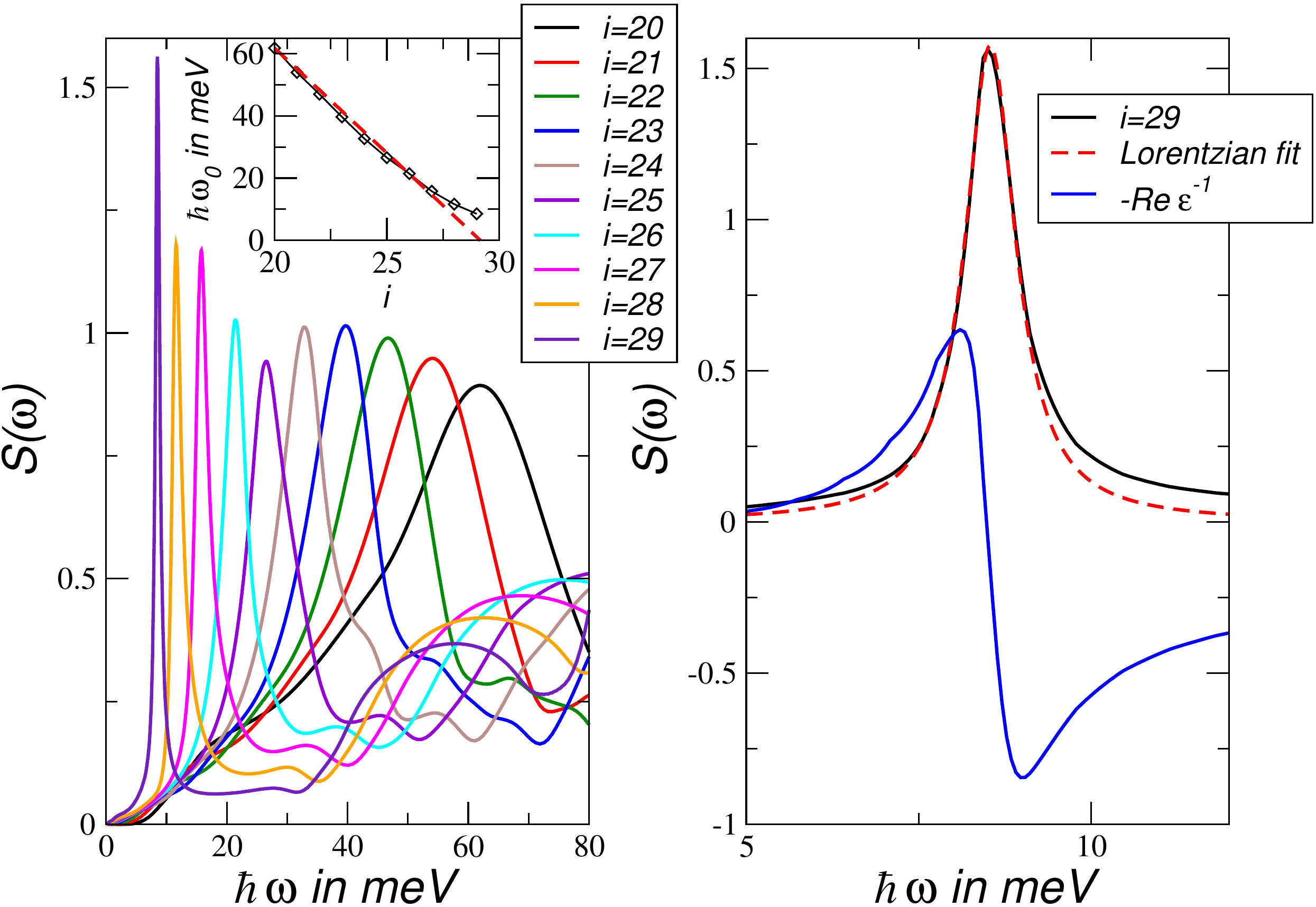}
\caption{Loss function $S(\omega)=$-Im$\epsilon^{-1}(\q,\omega)$ for $|\q|a=0.02$ in the $KK'$-direction for $\epsilon_0=4.8$ for twist angles with $i=20-29$ (left). For $i=29$, also a Lorentzian fit is shown with $\tilde S(\omega)=2C\gamma[(\omega-\omega_0)^2+\gamma^2]^{-1}$ where $\hbar C=3$meV, $\hbar\gamma=4$meV, and $\hbar\omega_0\sim8$meV.
\label{LossFunction}}
\end{figure}

\section{Entropy of the electron liquid}

The electronic contribution to the entropy $S$ can be obtained by applying the formula\cite{Abrikosov63}
\begin{align}
\frac{S}{A} = \frac{i}{\pi}\frac{1}{T} \int \frac{d^2 k}{(2\pi)^2} \int_{-\infty }^{\infty } d\omega \: \omega \frac{\partial n_F(\omega)}{\partial \omega } \log \left( \frac{G_R (\k, \omega )}{G_A (\k, \omega )} \right)\;,
\end{align}
where $A$ is the area of the system and $G_R , G_A$ are the retarded and advanced electron Green's functions, respectively. When looking for the low-temperature dependence of the entropy, one can perform the momentum integral along the Fermi line by decomposing $\k $ into longitudinal $k_{\parallel }$ and transverse $k_{\perp }$ components. This leads to 
\begin{align}
\frac{S}{A}  \approx   \frac{1}{\pi^2 T} \oint  \frac{dk_{\parallel }}{2\pi } \int \frac{d\varepsilon_{\k}}{v_{\k}} \int_{-\infty }^{\infty } d\omega \: \omega \frac{\partial n_F(\omega)}{\partial \omega }    \arctan \left( \frac{{\rm Im} \: \Sigma (\k, \omega )}{\omega - {\rm Re} \: \Sigma (\k, \omega ) - \varepsilon_{\k}}  \right) 
\end{align} 
where the integral in $k_{\parallel }$ is carried out along the Fermi line. 
The integral over the energy variable $\varepsilon_{\k}$ can be computed by adopting a principal value prescription. Then we get
\begin{align}
\frac{S}{A}  \approx \frac{1}{2 \pi^2} \frac{1}{T} \oint  \frac{dk_{\parallel }}{v_{\k}}  \int_{-\infty }^{\infty } d\omega \: \omega \frac{\partial n_F(\omega)}{\partial \omega }  \left( \omega - {\rm Re} \: \Sigma (\k, \omega ) \right)
\label{entr}
\end{align}

The temperature dependence of the entropy can be estimated by absorbing $T$ into a dimensionless variable $\omega/T$ in the integrand of Eq. (\ref{entr}). In particular, when the real part of the electron self-energy has an anomalous logarithmic correction, we see that this is translated to the entropy, which gets a dominant scaling behavior $S \sim T \: |\log(T)|$.

\end{widetext}
\end{document}